# Trapping of Noble Gases (He-Kr) by the Aromatic $H_3^+$ and $Li_3^+$ Species: A Conceptual DFT Approach

**Arindam Chakraborty, Santanab Giri and Pratim Kumar Chattaraj** *

*Department of Chemistry & Center for Theoretical Studies, Indian Institute of Technology, Kharagpur 721302, India*

E-mail: pkc@chem.iitkgp.ernet.in

## Abstract

Stability, reactivity and aromaticity of clusters of various noble gas atoms trapped in aromatic $H_3^+$ and $Li_3^+$ rings are studied at the B3LYP/6-311+G(d) and MP2/6-311+G(d) levels of theory. Electrophilicity, gain in energy and nucleus independent chemical shift values lend additional insights into the overall behavior of these clusters.

**Introduction:**

The noble gases occupy a dominant position in the periodic table of the elements. Although they were earlier thought to hardly have any reactivity and hence named inert, the jinx was broken for the first time by Bartlett[1] with the synthesis of $Xe^+[PtF_6]^-$. Since then the unexplored chemistry of the noble gases flourished in leaps and bounds and motivated both the synthetic and theoretical chemists to deliver their part[2]. The recent discovery of noble gas insertion compounds having vital terrestrial applications[2a,3] and experimental and theoretical studies of neutral van der Waals complexes of the noble gases with large and small molecular moieties[4] further nourished the chemical aspects of these so-called inert systems. Pauzat and co-workers[5] in their series of recent studies on the reactivity of the noble gases have discussed the prospect of $H_3^+$ as a trap for these rare gases. On the other hand it is well established that the seminal concept of aromaticity in conjugated organic systems is intensely related to their unusual stability compared to the open-chain analogs. Moreover the existence of "all-metal aromaticity" in the purely inorganic $Al_4^{2-}$ cluster proposed by Boldyrev et al[6] further strengthened the splendid correlation between the phenomenon of aromaticity and stability of metallic as well as non-metallic clusters. The $H_3^+$ unit chosen by Pauzat et al[5] as a catch for the noble gases along with its another group compatriot, $Li_3^+$ have recently been reported to possess σ-aromaticity in terms of ring current and NICS[7] measures. The $Li_3^+$ unit, however, does not show any ring current[8] but leads to a negative NICS value, the essential aromaticity criterion.

In this work we intend to investigate the potential ability of the $H_3^+$ as well as $Li_3^+$ units as possible traps for the noble gases (He-Kr) from the viewpoint of conceptual density functional theory and aromaticity. The stability of these trapped noble-gas complexes can be well elucidated under the paradigm of conceptual DFT[9] in association with its various global reactivity descriptors like electronegativity[10] (χ), hardness[11] (η), electrophilicity[12] (ω) and the local variants like atomic charges[13] ($Q_k$) and Fukui functions[14] ($f_k$). The aromaticity criterion for the $H_3^+$ and $Li_3^+$ trigonal rings in the noble-gas complexes measured in terms of the NICS[7] shall convey valuable insights into judging the stability of the noble gas complexes.

**Theoretical Background:**

Electrophilicity (ω) does play a vital role in quantifying the thermodynamic stability and reactivity of molecular systems. A minimum electrophilicity Principle[15] (MEP) has been proposed for this purpose. In an N-electron system, the electronegativity[10] (χ) and hardness[11] (η) can be defined as follows:

$$\chi = -\mu = -\left(\frac{\partial E}{\partial N}\right)_{v(\vec{r})} \qquad (1)$$

$$\eta = \left(\frac{\partial^2 E}{\partial N^2}\right)_{v(\vec{r})} \qquad (2)$$

Here $E$ is the total energy of the $N$-electron system and μ and $v(\vec{r})$ are its chemical potential and external potential respectively. The electrophilicity[12] (ω) is defined as:

$$\omega = \frac{\mu^2}{2\eta} = \frac{\chi^2}{2\eta} \qquad (3)$$

A finite difference approximation to Eqs. 1 and 2 can be expressed as:

$$\chi = \frac{I + A}{2} \tag{4}$$

and $$\eta = I - A \tag{5}$$

where $I$ and $A$ represent the ionization potential and electron affinity of the system respectively and are computed in terms of the energies of the $N$ and $N \pm 1$ electron systems. For an $N$-electron system with energy $E\ (N)$ they may be expressed as follows:

$$I = E\ (N-1) - E(N) \tag{6}$$

and $$A = E(N) - E(N+1) \tag{7}$$

The local reactivity descriptor, Fukui function[14] (FF) measures the change in electron density at a given point when an electron is added to or removed from a system at constant $v(\vec{r})$. It may be written as:

$$f(\vec{r}) = \left( \frac{\partial \rho(\vec{r})}{\partial N} \right)_{v(\vec{r})} = \left( \frac{\delta \mu}{\delta v(\vec{r})} \right)_N \tag{8}$$

Condensation of this Fukui function, $f(\vec{r})$ to an individual atomic site $k$ in a molecule gives rise to the following expressions in terms of electron population[16] $q_k$

$f_k^+ = q_k(N+1) - q_k(N)$ for nucleophilic attack (9a)

$f_k^- = q_k(N) - q_k(N-1)$ for electrophilic attack (9b)

$f_k^o = [q_k(N+1) - q_k(N-1)]/2$ for radical attack (9c)

**Computational Details:**

The geometry optimization of the molecular conformations of the trapped noble-gas clusters and their subsequent frequency calculations are carried out at the B3LYP and MP2 levels of theory using the 6-311+G(d) molecular basis set with the aid of GAUSSIAN 03 program package[17]. The NIMAG values of all the optimized geometries are zero thereby confirming their existence at the minima on the potential energy surface (PES). Single point calculations are further done to evaluate the energies of the $N \pm 1$ electron systems by adopting the geometries of the corresponding $N$-electron systems optimized at the B3LYP/6-311+G(d) level of theory. The $I$ and $A$ values are calculated using a $\Delta SCF$ technique. The electrophilicity (ω) and hardness (η) were computed using the eqs. 3 and 5 respectively. A Mulliken population analysis (MPA) scheme is adopted to calculate the atomic charges ($Q_k$) and Fukui functions ( $f(\bar{r})$ ). The NICS[7] values at the center (NICS(0)) of the trigonal $H_3^+$ and $Li_3^+$ rings as well as at different distances perpendicular to the ring center are calculated. The frontier molecular orbital pictures are obtained through the GAUSSVIEW 03 package[17].

**Results and Discussion:**

The global reactivity descriptors like electronegativity (χ), hardness (η) and electrophilicity (ω) values of the four noble gas atoms (He-Kr), $H_3^+$ and its trapped clusters and that of $Li_3^+$ and its corresponding trapped cluster molecules are presented in tables 1 and 2 respectively. The molecular point groups (PG) and the NICS values at different distances from the ring center for the $H_3^+$ and $Li_3^+$ trapped noble gas clusters are put forward in tables 3 and 4 respectively. The energy (E) values of the $H_3^+$ and $Li_3^+$ trapped noble gas clusters computed at different levels of theory using the 6-311+G(d) basis set are presented in tables 5 and 6 respectively. A detailed population analysis under the Mulliken scheme (MPA) consisting of the atomic charges ($Q_k$) and fukui functions ($f_k^+$, $f_k^-$) for all the atomic sites of the $H_3^+$ and $Li_3^+$ trapped noble gas clusters are shown in tables 7 and 8 respectively. Tables 9 and 10 depict some plausible complexation reactions that may occur during the trapping of the noble gas atoms by $H_3^+$ or $Li_3^+$ respectively. The feasibility of such reactions in real practice may be justified from their ΔH or $\Delta\omega$ values. Figures 1 and 2 depict the stable molecular conformations of the $H_3^+$ and $Li_3^+$ trapped noble gas clusters respectively. Figures 3 (a-d) and 4 (a-c) illustrate the gain in energy ($\Delta E$) computed at B3LYP level and electrophilicity ($\omega$) of the different noble gas atomic assemblies trapped by the $H_3^+$ and $Li_3^+$ aromatic systems. Figures 5 and 6 portray the important frontier molecular orbitals (FMOs) of the $H_3^+$ and $Li_3^+$ trapped noble gas clusters respectively. From tables 1 and 2 it transpires that $H_3^+$ can serve as a better trap for the noble gas atoms as it has got the ability to hold any of them amongst He-Kr. For $Ar_nH_3^+$ cluster as many as five Ar atoms can be bound together by the $H_3^+$ trigonal ring. The situation is nevertheless not so encouraging with the $Li_3^+$ unit as it can trap He, Ne and Ar atoms and not Kr. The hardness ($\eta$) and electrophilicity ($\omega$) for the noble gas atoms follow the expected pattern, with the gradual increment in the atomic size from He to Kr, the hardness ($\eta$) falls as the atoms get softer and thereby encouraging an increment in the corresponding electrophilicity ($\omega$) values. Thus the noble gases are supposed to be more reactive as one move down the periodic group from He to Kr. However, the hardness ($\eta$) and electrophilicity ($\omega$) trends show some interesting outcomes for the trapped noble atomic clusters. For the $Li_3^+$ trapped clusters it is observed from table 2 that the $\eta$ values more or less increase with increasing cluster size (n value) for corresponding noble gas atoms which is accompanied with a gradual decrease in their respective $\omega$ values. This signifies a sheer unwillingness of the larger $Li_3^+$-trapped noble atomic clusters towards chemical reactivity, a phenomenon in accord with the MEP criterion. Of course both the chemical and external potentials change drastically. Moreover, it is expected that a system would get softer with an increase in its size. The situation is, however, not so much straight-forward for the corresponding $H_3^+$-trapped noble gas clusters. Table 1 shows that for all the noble gas atoms, the $\eta$ values decrease with increasing cluster size which is a trademark of an increase in chemical reactivity. On the contrary, the $\omega$ values also show a hand-in-hand decreasing trend! From tables 3 and 4, it becomes relevant that both the $H_3^+$ and $Li_3^+$ rings in their free uncombined as well as in trapped condition show a highly negative NICS(0) value, a phenomenon pointing towards high stability of the molecular systems in terms of aromaticity[7]. It can perhaps now be assured that although the $\eta$ and $\omega$ trends for the $H_3^+$ trapped noble gas clusters show some anomaly, the stability of the $H_3^+$ unit upon trapping the noble gas atoms, as many as five for Ar can well be settled from the viewpoint of aromaticity criterion in terms of NICS measures. It is further noticed that the NICS values show an obvious decrease as one move away from the ring centers. The situation becomes a bit exciting for the $Ar_5H_3^+$ cluster where NICS(3) and NICS(4) values are much negative than the NICS(0) value. This trend becomes justified upon consideration of the trigonal bi-pyramidal arrangement of the five Ar atoms around the triangular $H_3^+$ unit which, of course, deserves a careful scrutiny. The ground state energies of all the noble gas trapped stable molecular clusters of $H_3^+$ and $Li_3^+$ computed at different levels of theory are shown in tables 5 and 6 respectively. Qualitative trends remain same in both the levels. A close scrutiny of table 7 reveals that the variation of the Ng–H (Ng = Ne, Ar, Kr) bond distances for the different noble-gas trapped clusters follow the same trend as reported by Pauzat et.al[5]. Moreover, it may be noted that the most stable structure among the several stationary points for the trapped-clusters reported by Pauzat et al[5] exactly correspond to the ones obtained in this present study. From the information regarding the local parameters presented in tables 8 and 9 it is observed that the atomic charges ($Q_k$) for almost all the atomic sites of the $H_3^+$ and $Li_3^+$ trapped clusters are positive. Although the $Q_k$ values in all three atomic sites of $H_3^+/Li_3^+$ are same due to symmetry, the same behavior is not observed in the corresponding $f_k$ values as the uniformity in $Q_k$ values is lost in the cases of the respective anions (i.e., neutral $H_3/Li_3$) calculated using a single point method on the geometry of the corresponding N- electron species ($H_3^+/Li_3^+$). This presupposes a nucleophilic or radical attack at all the atomic centers. The gradual acceptance of neutral noble gas atoms by the $H_3^+$ and $Li_3^+$ units to form larger stable clusters further strengthens this point. A scrutiny of tables 10 and 11 reveals that the reaction enthalpy (ΔH) for the first-phase trapping reactions between $H_3^+$ and/or $Li_3^+$ systems with a single noble gas atom becomes more favorable as one moves from He to Kr, thereby lending ample justice to the increasing reactivities of the noble atoms downward a periodic group. Thus the first-phase trapping reactions among $H_3^+$ and/or $Li_3^+$ with the noble gases are thermodynamically quite favorable. However, for the second-phase and further higher order trapping reactions, the ΔH values relatively increase for the corresponding noble gas atoms. This may be attributed to the phenomenon of "steric crowding" which probably comes into play with an increase in the number of atoms around the trigonal $H_3^+$ or $Li_3^+$ ring. The reaction electrophilicity (Δω) values for the trapping reactions with the $H_3^+$ moiety relatively follow the same pattern as dictated by their corresponding ΔH values. But for the $Li_3^+$-trapping reactions, the Δω values markedly decrease with further trapping thereby encouraging the process further. Thus it may be inferred that although the $H_3^+$ or $Li_3^+$ units show the ability to trap a number of noble gas atoms, the corresponding trapped clusters, particularly those with heavier atoms may exhibit kinetic stability and are of "fleeting" type[18]. Figures 3 (a-d) pictorially demonstrate the variation of gain in energy (ΔE) and electrophilicity (ω) values for the $H_3^+$-trapped noble atomic clusters. Both of these descriptors show a decreasing trend upon gradual cluster growth of $H_3^+$ with the noble atoms, as expected from the tables 10 and 11. The gain in energy (ΔE) values show a consistent falling trend except for the pairs $Ar_4H_3^+$ – $Ar_5H_3^+$, $Kr_2H_3^+$ – $Kr_3H_3^+$ and $Ne_2Li_3^+$ – $Ne_3Li_3^+$. Among the former $H_3^+$-trapped clusters, the values are quite comparable whereas for the latter $Li_3^+$-trapped pair, the ΔE value shows a reverse trend. This may be attributed to a relatively lesser increase in the energy (E) of the system during higher order trapping. It is also quite evident from tables 5 and 6 that the energy difference between the $Ne_2Li_3^+$ and $Ne_3Li_3^+$ clusters is marginally higher compared to the other two $H_3^+$-trapped noble gas cluster pairs. Figures 4 (a-c) describe the trends of the energy gain (ΔE) and electrophilicity (ω) values upon trapping of $Li_3^+$ with He, Ne and Ar respectively. As discussed earlier, the $Li_3^+$-bound noble gas clusters are found to obey the minimum electrophilicity principle (MEP). As shown in figures 5 and 6 essential σ-symmetry in the frontier molecular orbitals of $H_3^+$ and $Li_3^+$ trigonal rings are retained in most of the $Li_3^+$- trapped noble atomic clusters whereas for the $H_3^+$- trapped clusters additional π- symmetry is introduced.

**Concluding Remarks:**

A study of the trapping of the noble gas atoms (He-Kr) by the trigonal $H_3^+$ and $Li_3^+$ systems within a conceptual density functional theory framework reveals that a considerable amount of stability can be guaranteed for the so-called inert gas atoms upon forming small to medium sized cationic clusters. This stability can further be justified by the aromaticity criterion defined in terms of NICS measures. Further efforts to rationalize the stability of the $H_3^+$ and $Li_3^+$ bound noble atomic clusters from the kinetic as well as thermodynamic perspectives are also attempted from a study of the reaction enthalpy ($\Delta H$) and reaction electrophilicity ($\Delta\omega$) values of the plausible step-wise trapping reactions. In a nutshell, it may be outlined that unlike the first-phase trapping steps, second and higher-phase trapping processes may be marginally favorable. The stability of the medium-sized trapped clusters may be a bit at stake due to steric crowding.

**Acknowledgement:**

We are thankful to the Indo-EU project HYPOMAP for financial assistance.

**Table 1**: The electronegativity (χ), hardness (η), and electrophilicity (ω) values for the noble gas atoms and the corresponding $H_3^+$-trapped clusters.

| Molecule | χ(eV) | η(eV) | ω(eV) |
|---|---|---|---|
| **He** | 1.770 | 46.310 | 0.034 |
| **Ne** | 7.609 | 28.384 | 1.020 |
| **Ar** | 6.450 | 18.800 | 1.106 |
| **Kr** | 6.216 | 16.008 | 1.207 |
| **$H_3^+$** | 19.928 | 27.380 | 7.252 |
| **$HeH_3^+$** | 18.126 | 24.060 | 6.828 |
| **$He_2H_3^+$** | 17.085 | 22.624 | 6.451 |
| **$He_3H_3^+$** | 16.354 | 22.218 | 6.018 |
| **$NeH_3^+$** | 16.918 | 21.962 | 6.516 |
| **$Ne_2H_3^+$** | 15.641 | 20.144 | 6.072 |
| **$Ne_3H_3^+$** | 14.912 | 19.652 | 5.658 |
| **$ArH_3^+$** | 13.985 | 17.530 | 5.578 |
| **$Ar_2H_3^+$** | 12.520 | 15.394 | 5.091 |
| **$Ar_3H_3^+$** | 11.781 | 14.828 | 4.680 |
| **$Ar_4H_3^+$** | 11.393 | 14.088 | 4.606 |
| **$Ar_5H_3^+$** | 11.079 | 13.510 | 4.543 |
| **$KrH_3^+$** | 13.389 | 16.144 | 5.552 |
| **$Kr_2H_3^+$** | 11.707 | 13.552 | 5.056 |
| **$Kr_3H_3^+$** | 10.785 | 13.286 | 4.377 |

**Table 2**: The electronegativity (χ), hardness (η), and electrophilicity (ω) values for the $Li_3^+$-trapped clusters of noble gas atom.

| Molecule | Electronegativity (χ) | Hardness (η) | Electrophilicity (ω) |
|---|---|---|---|
| **$Li_3^+$** | 7.345 | 6.444 | 4.186 |
| **$HeLi_3^+$** | 7.306 | 6.424 | 4.154 |
| **$He_2Li_3^+$** | 7.261 | 6.502 | 4.054 |
| **$He_3Li_3^+$** | 7.111 | 6.616 | 3.821 |
| **$NeLi_3^+$** | 7.290 | 6.422 | 4.136 |
| **$Ne_2Li_3^+$** | 7.209 | 6.448 | 4.029 |
| **$Ne_3Li_3^+$** | 7.093 | 6.562 | 3.833 |
| **$ArLi_3^+$** | 7.192 | 6.356 | 4.069 |
| **$Ar_2Li_3^+$** | 7.028 | 6.316 | 3.910 |
| **$Ar_3Li_3^+$** | 6.793 | 6.396 | 3.608 |

**Table 3**: The molecular point group (PG) and NICS values of $H_3^+$ and corresponding trapped noble gas clusters.

| **Molecule** | **Point Group (PG)** | **NICS(0) (ppm)** | **NICS(1) (ppm)** | **NICS(2) (ppm)** | **NICS(3) (ppm)** | **NICS(4) (ppm)** |
|---|---|---|---|---|---|---|
| **$H_3^+$** | $D_{3h}$ | - 30.66 | - 1.75 | - 0.03 | - 0.01 | - 0.00 |
| **$HeH_3^+$** | $C_{2v}$ | - 30.58 | - 1.81 | - 0.07 | - 0.02 | - 0.01 |
| **$He_2H_3^+$** | $C_{2v}$ | - 30.56 | - 1.86 | - 0.10 | - 0.04 | - 0.02 |
| **$He_3H_3^+$** | $C_s$ | - 30.56 | - 1.91 | - 0.13 | - 0.06 | - 0.03 |
| **$NeH_3^+$** | $C_{2v}$ | - 30.19 | - 1.84 | - 0.08 | - 0.03 | - 0.01 |
| **$Ne_2H_3^+$** | $C_{2v}$ | - 29.94 | - 1.90 | - 0.12 | - 0.05 | - 0.02 |
| **$Ne_3H_3^+$** | $C_s$ | - 29.74 | - 1.91 | - 0.15 | - 0.06 | - 0.03 |
| **$ArH_3^+$** | $C_{2v}$ | - 24.79 | - 1.44 | - 0.05 | - 0.03 | - 0.01 |
| **$Ar_2H_3^+$** | $C_{2v}$ | -25.39 | - 1.39 | - 0.12 | - 0.07 | - 0.04 |
| **$Ar_3H_3^+$** | $C_1$ | - 25.55 | - 1.40 | - 0.19 | - 0.11 | - 0.06 |
| **$Ar_4H_3^+$** | $C_1$ | - 25.46 | - 1.38 (below the plane) | - 0.18 (below the plane) | - 0.11 (below the plane) | -0.06 (below the plane) |
| **$Ar_5H_3^+$** | $C_1$ | - 25.37 | - 1.30 | - 4.71 | - 88.36 | - 41.91 |
| **$KrH_3^+$** | $C_s$ | -20.28 | -1.24 | 0.01 | 0.00 | 0.00 |
| **$Kr_2H_3^+$** | $C_{2v}$ | -22.16 | -0.93 | -0.02 | -0.03 | -0.01 |
| **$Kr_3H_3^+$** | $C_1$ | -22.82 | -0.87 | -0.09 | -0.07 | -0.04 |

**Table 4:** The molecular point group (PG) and NICS values of $Li_3^+$ and corresponding trapped noble gas clusters.

| **Molecule** | **Point Group (PG)** | **NICS(0) (ppm)** | **NICS(1) (ppm)** | **NICS(2) (ppm)** | **NICS(3) (ppm)** | **NICS(4) (ppm)** |
|---|---|---|---|---|---|---|
| **$Li_3^+$** | $D_{3h}$ | -11.08 | -6.78 | -1.61 | -0.10 | 0.04 |
| **$HeLi_3^+$** | $C_{2v}$ | -11.08 | -6.79 | -1.64 | -0.12 | 0.03 |
| **$He_2Li_3^+$** | $C_{2v}$ | -11.09 | -6.80 | -1.66 | -0.13 | 0.02 |
| **$He_3Li_3^+$** | $C_1$ | -11.05 | -6.81 | -1.71 | -0.16 | 0.02 |
| **$NeLi_3^+$** | $C_{2v}$ | -11.11 | -6.81 | -1.64 | -0.12 | 0.03 |
| **$Ne_2Li_3^+$** | $C_{2v}$ | -11.14 | -6.83 | -1.68 | -0.14 | 0.02 |
| **$Ne_3Li_3^+$** | $C_1$ | -11.17 | -6.86 | -1.71 | -0.16 | 0.01 |
| **$ArLi_3^+$** | $C_{2v}$ | -11.12 | -6.83 | -1.69 | -0.16 | 0.01 |
| **$Ar_2Li_3^+$** | $C_{2v}$ | -11.15 | -6.88 | -1.77 | -0.21 | -0.02 |
| **$Ar_3Li_3^+$** | $C_1$ | -11.16 | -6.92 | -1.86 | -0.27 | -0.05 |

**Table 5**: Energy (E, Hartrees) of $H_3^+$ and corresponding trapped noble gas clusters computed at different levels of theory using 6-311+G(d) basis set.

| Molecule | Energy (B3LYP/6-311+G(d)) | Energy (MP2/6-311+G(d)) |
|---|---|---|
| **$H_3^+$** | -1.32900 | -1.29898 |
| **$HeH_3^+$** | -4.24386 | -4.17223 |
| **$He_2H_3^+$** | -7.15859 | -7.04548 |
| **$He_3H_3^+$** | -10.07323 | -9.91873 |
| **$NeH_3^+$** | -130.29288 | -130.03971 |
| **$Ne_2H_3^+$** | -259.25631 | -258.78046 |
| **$Ne_3H_3^+$** | -388.21957 | -387.52132 |
| **$ArH_3^+$** | -528.89794 | -528.26072 |
| **$Ar_2H_3^+$** | -1056.45796 | -1055.22070 |
| **$Ar_3H_3^+$** | -1584.01688 | -1582.18042 |
| **$Ar_4H_3^+$** | -2111.57152 | -2109.13746 |
| **$Ar_5H_3^+$** | -2639.12623 | -2636.09467 |
| **$KrH_3^+$** | -2755.11007 | -2753.40743 |
| **$Kr_2H_3^+$** | -5508.87013 | -5505.50934 |
| **$Kr_3H_3^+$** | -8262.62975 | -8257.61128 |

**Table 6:** Energy (E, Hartrees) of $Li_3^+$ and corresponding trapped noble gas clusters computed at different levels of theory using 6-311+G(d) basis set.

| Molecule | Energy (B3LYP/6-311+G(d)) | Energy (MP2/6-311+G(d)) |
|---|---|---|
| **$Li_3^+$** | -22.37210 | -22.19253 |
| **$HeLi_3^+$** | -25.28604 | -25.06556 |
| **$He_2Li_3^+$** | -28.19984 | -27.93858 |
| **$He_3Li_3^+$** | -31.11363 | -30.8116 |
| **$NeLi_3^+$** | -151.33463 | -150.93399 |
| **$Ne_2Li_3^+$** | -280.29710 | -279.67540 |
| **$Ne_3Li_3^+$** | -409.25959 | -408.41681 |
| **$ArLi_3^+$** | -549.92982 | -549.15350 |
| **$Ar_2Li_3^+$** | -1077.48722 | -1076.11422 |
| **$Ar_3Li_3^+$** | -1605.04448 | -1603.07465 |

| **Molecule** | **Ng–H bond length (Å) [Ng = Ne, Ar, Kr]** | **Pauzat's results* [BH&HLYP/cc-pVTZ]** | **Present study [B3LYP/6-311+G(d)]** |
|---|---|---|---|
| **$NeH_3^+$** | Ne–H | 1.723 | 1.769 |
| **$Ne_2H_3^+$** | Ne–H | 1.778 | 1.819 |
| **$Ne_3H_3^+$** | Ne–H | 1.817 | 1.855 |
| **$ArH_3^+$** | Ar–H | 1.791 | 1.709 |
| **$Ar_2H_3^+$** | Ar–H | 1.984 | 1.969 |
| **$Ar_3H_3^+$** | Ar–H | 2.082 | 2.088 |
| **$Ar_4H_3^+$** | Ar–H | 2.087 | 2.096 |
| **$Ar_5H_3^+$** | Ar–H | 2.094 | 2.098 |
| **$KrH_3^+$** | Kr–H | 1.799 | 1.665 |
| **$Kr_2H_3^+$** | Kr–H | 2.093 | 2.026 |
| **$Kr_3H_3^+$** | Kr–H | 2.211 | 2.190 |

**Table 7**: Comparison of the Ng–H (Ng = Ne, Ar, Kr) bond lengths (Å) obtained from Pauzat's calculations (Reference 5(a-c)) and the present study.
*values are taken from references 5(a-c)

**Table 8**: The atomic charges ($Q_k$) and fukui functions ($f_k^+$, $f_k^-$) of $H_3^+$ and corresponding trapped noble gas clusters computed under the MPA formulation.

| **Molecule** | **Unit** | **$Q_k$** | **$f_k^+$** | **$f_k^-$** |
|---|---|---|---|---|
| **$H_3^+$** | H, H, H | 0.333, 0.333, 0.333 | 0.077, 0.462, 0.462 | 0.333, 0.333, 0.333 |
| **$HeH_3^+$** | H, H, H<br>He | 0.296, 0.331, 0.331,<br>0.042 | 0.059, 0.462, 0.461,<br>0.018 | 0.027, 0.187, 0.187,<br>0.598 |
| **$He_2H_3^+$** | H, H, H<br>He, He | 0.296, 0.329, 0.296,<br>0.040, 0.040 | 0.149, 0.681, 0.142,<br>0.014, 0.014 | 0.025, 0.130, 0.026,<br>0.467, 0.351 |
| **$He_3H_3^+$** | H, H, H<br>He, He, He | 0.295, 0.295, 0.295,<br>0.038, 0.038, 0.038 | 0.220, 0.161, 0.593,<br>0.012, 0.000, 0.014 | 0.041, 0.041, 0.041,<br>0.292, 0.292, 0.292 |
| **$NeH_3^+$** | H, H, H<br>Ne | 0.268, 0.331, 0.331,<br>0.069 | 0.066, 0.459, 0.443,<br>0.032 | -0.062, 0.138, 0.139,<br>0.785 |
| **$Ne_2H_3^+$** | H, H, H<br>Ne, Ne | 0.272, 0.331, 0.272,<br>0.062, 0.062 | 0.155, 0.675, 0.152,<br>0.009, 0.009 | -0.007, 0.106, -0.007,<br>0.453, 0.454 |
| **$Ne_3H_3^+$** | H, H, H<br>Ne, Ne, Ne | 0.273, 0.274, 0.273,<br>0.060, 0.060, 0.060 | 0.267, 0.144, 0.663,<br>-0.081, -0.007, 0.014 | 0.022, 0.022, 0.021,<br>0.310, 0.311, 0.312 |
| **$ArH_3^+$** | H, H, H<br>Ar | 0.247, 0.251, 0.251,<br>0.250 | 0.443, 0.216, 0.222,<br>0.118 | -0.056, 0.091, 0.091,<br>0.875 |
| **$Ar_2H_3^+$** | H, H, H<br>Ar, Ar | 0.222, 0.257, 0.222,<br>0.150, 0.150 | 0.401, 0.029, 0.401,<br>0.084, 0.084 | -0.002, 0.072, -0.002,<br>0.466, 0.466 |
| **$Ar_3H_3^+$** | H, H, H | 0.214, 0.214, 0.213, | 0.203, 0.055, 0.483, | 0.006, 0.005, 0.006, |
| | Ar, Ar, Ar | 0.121, 0.117, 0.120 | 0.063, 0.103, 0.093 | 0.324, 0.333, 0.326 |
| | H, H, H | 0.213, 0.212, 0.213, | -0.004, 0.430, 0.315, | -0.008, -0.008, -0.003, |
| | Ar, Ar, Ar, | -0.001, 0.120, 0.121, | 0.014, 0.106, 0.062, | 0.439, 0.195, 0.191, |
| **$Ar_4H_3^+$** | Ar | 0.120 | 0.077 | 0.194 |
| | H, H, H | 0.212, 0.212, 0.212, | -0.014, 0.401, 0.353, | -0.015, -0.015, -0.015, |
| | Ar, Ar, Ar, | -0.001, -0.001, 0.122, | 0.014, 0.013, 0.105, | 0.376, 0.383, 0.097, |
| **$Ar_5H_3^+$** | Ar, Ar | 0.122, 0.122 | 0.061, 0.067 | 0.101, 0.089 |
| | H, H, H | 0.238, 0.190, 0.190, | 0.416, 0.147, 0.141, | -0.024, 0.085, 0.086, |
| **$KrH_3^+$** | Kr | 0.382 | 0.295 | 0.852 |
| | H, H, H | 0.200, 0.216, 0.200, | 0.336, 0.030, 0.336, | 0.000, 0.066, 0.000, |
| **$Kr_2H_3^+$** | Kr, Kr | 0.192, 0.192 | 0.148, 0.148 | 0.466, 0.468 |
| | H, H, H | 0.193, 0.193, 0.193, | 0.081, 0.130, 0.433, | 0.000, 0.000, 0.000, |
| **$Kr_3H_3^+$** | Kr, Kr, Kr | 0.142, 0.139, 0.140 | 0.117, 0.120, 0.119 | 0.328, 0.337, 0.334 |

**Table 9**: The atomic charges ($Q_k$) and fukui functions ($f_k^+$, $f_k^-$) of $Li_3^+$ and corresponding trapped noble gas clusters computed under the MPA formulation.

| Molecule | Unit | $Q_K$ | $f_K^+$ | $f_K^-$ |
|---|---|---|---|---|
| **$Li_3^+$** | Li, Li, Li | 0.333, 0.333, 0.333 | 0.209, 0.395, 0.395 | 0.333, 0.333, 0.333 |
| **$HeLi_3^+$** | Li, Li, Li | 0.369, 0.215, 0.374 | 0.341, 0.331, 0.322 | 0.305, 0.386, 0.299 |
| | He | 0.041 | 0.005 | 0.010 |
| **$He_2Li_3^+$** | Li, Li, Li | 0.258, 0.262, 0.407 | 0.303, 0.304, 0.384 | 0.354, 0.351, 0.276 |
| | He, He | 0.036, 0.037 | 0.004, 0.005 | 0.009, 0.009 |
| **$He_3Li_3^+$** | Li, Li, Li | 0.295, 0.297, 0.282 | 0.433, 0.104, 0.449 | 0.320, 0.313, 0.333 |
| | He, He, He | 0.042, 0.042, 0.043 | 0.006, 0.004, 0.004 | 0.011, 0.011, 0.011 |
| **$NeLi_3^+$** | Li, Li, Li | 0.329, 0.276, 0.329 | 0.406, 0.200,0.390 | 0.352, 0.279, 0.354 |
| | Ne | 0.066 | 0.003 | 0.015 |
| **$Ne_2Li_3^+$** | Li, Li, Li | 0.274, 0.273, 0.324 | 0.283, 0.240, 0.473 | 0.299, 0.300, 0.372 |
| | Ne, Ne | 0.064, 0.065 | 0.002, 0.002 | 0.014, 0.015 |
| **$Ne_3Li_3^+$** | Li, Li, Li | 0.267, 0.267, 0.267 | 0.444, 0.297, 0.263 | 0.318, 0.319, 0.319 |
| | Ne, Ne, Ne | 0.066, 0.066, 0.066 | 0.000, -0.010, 0.001 | 0.015, 0.015, 0.015 |
| **$ArLi_3^+$** | Li, Li, Li | 0.316, 0.248, 0.316 | 0.427, 0.146, 0.413 | 0.359, 0.237, 0.355 |
| | Ar | 0.119 | 0.014 | 0.049 |
| **$Ar_2Li_3^+$** | Li, Li, Li | 0.239, 0.240, 0.306 | 0.234, 0.229, 0.514 | 0.262, 0.260, 0.378 |
| | Ar, Ar | 0.108, 0.108 | 0.012, 0.012 | 0.050, 0.050 |
| **$Ar_3Li_3^+$** | Li, Li, Li | 0.225, 0.225, 0.225 | 0.358, 0.356, 0.296 | 0.283, 0.282, 0.280 |
| | Ar, Ar, Ar | 0.108, 0.108, 0.109 | 0.010, -0.010, -0.010 | 0.051, 0.051, 0.051 |

**Table 10**: Some plausible reactions that occur due to the attack of the noble gas atoms onto the $H_3^+$ moiety.

| No | Reactions | ΔH | Δω |
|---|---|---|---|
| 1 | **$H_3^+ + He = HeH_3^+$** | -0.445 | -0.917 |
| 2 | **$HeH_3^+ + He = He_2H_3^+$** | -0.541 | -0.820 |
| 3 | **$He_2H_3^+ + He = He_3H_3^+$** | -0.520 | -0.933 |
| 4 | **$H_3^+ + Ne = NeH_3^+$** | -1.488 | -3.511 |
| 5 | **$NeH_3^+ + Ne = Ne_2H_3^+$** | -1.389 | -2.927 |
| 6 | **$Ne_2H_3^+ + Ne = Ne_3H_3^+$** | -1.279 | -2.870 |
| 7 | **$H_3^+ + Ar = ArH_3^+$** | -9.570 | -2.870 |
| 8 | **$ArH_3^+ + Ar = Ar_2H_3^+$** | -3.027 | -3.188 |
| 9 | $Ar_2H_3^+ + Ar = Ar_3H_3^+$ | -2.497 | -3.035 |
| 10 | $Ar_3H_3^+ + Ar = Ar_4H_3^+$ | -0.240 | -2.360 |
| 11 | $Ar_4H_3^+ + Ar = Ar_5H_3^+$ | -0.223 | -2.340 |
| 12 | $H_3^+ + Kr = KrH_3^+$ | -17.866 | -5.813 |
| 13 | $KrH_3^+ + Kr = Kr_2H_3^+$ | -2.946 | -3.406 |
| 14 | $Kr_2H_3^+ + Kr = Kr_3H_3^+$ | -2.618 | -3.773 |

**Table 11**: Some plausible reactions that occur due to the attack of the noble gas atoms onto the $Li_3^+$ moiety.

| No | Reactions | ΔH | Δω |
|---|---|---|---|
| 1 | $Li_3^+ + He = HeLi_3^+$ | -0.247 | -0.131 |
| 2 | $HeLi_3^+ + He = He_2Li_3^+$ | -0.187 | -0.268 |
| 3 | $He_2Li_3^+ + He = He_3Li_3^+$ | -0.111 | -0.532 |
| 4 | $Li_3^+ + Ne = NeLi_3^+$ | -1.008 | -2.138 |
| 5 | $NeLi_3^+ + Ne = Ne_2Li_3^+$ | -0.982 | -2.254 |
| 6 | $Ne_2Li_3^+ + Ne = Ne_3Li_3^+$ | -0.982 | -2.432 |
| 7 | $Li_3^+ + Ar = ArLi_3^+$ | -2.070 | -2.446 |
| 8 | $ArLi_3^+ + Ar = Ar_2Li_3^+$ | -1.877 | -2.531 |
| 9 | $Ar_2Li_3^+ + Ar = Ar_3Li_3^+$ | -1.782 | -2.817 |

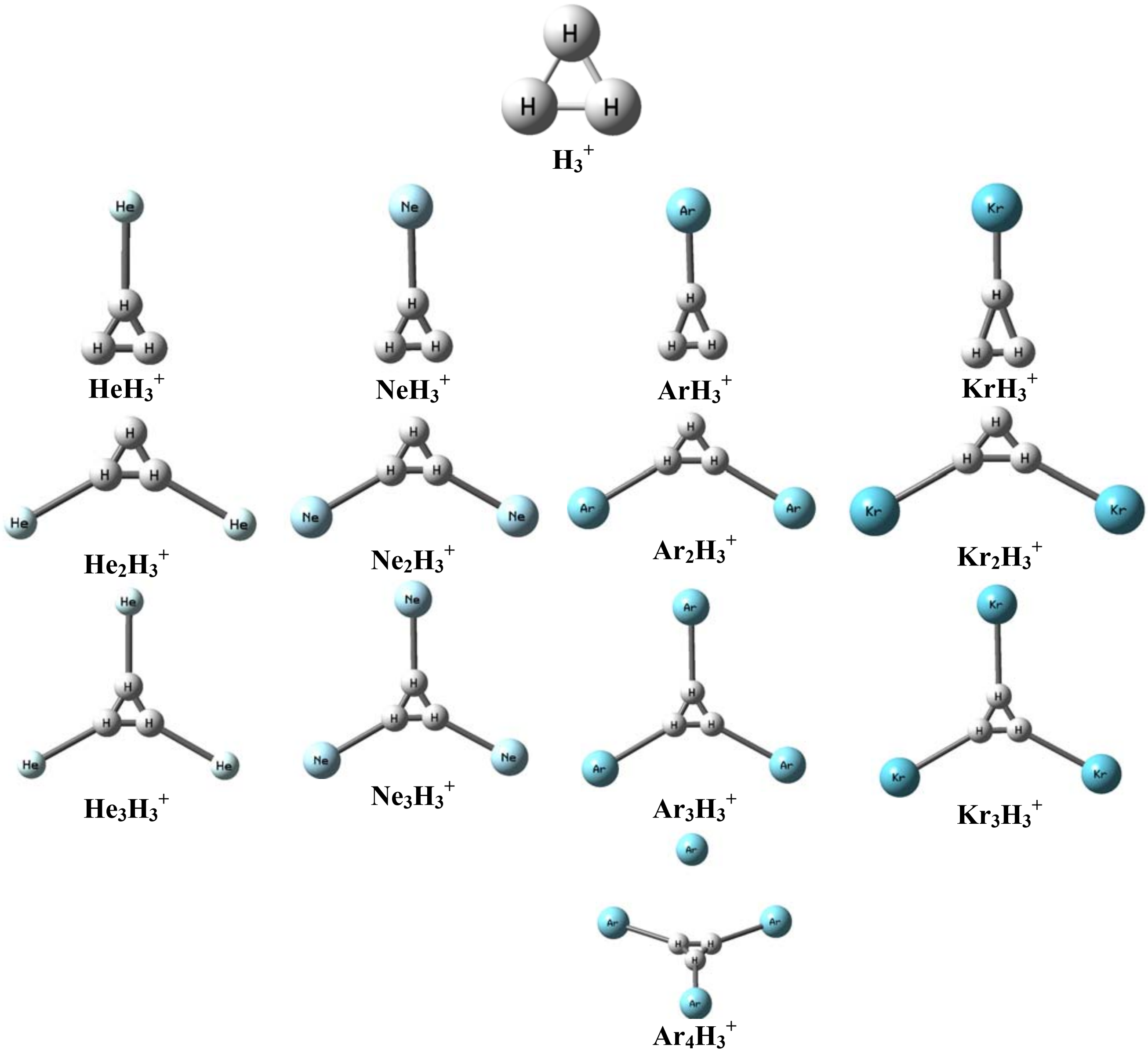

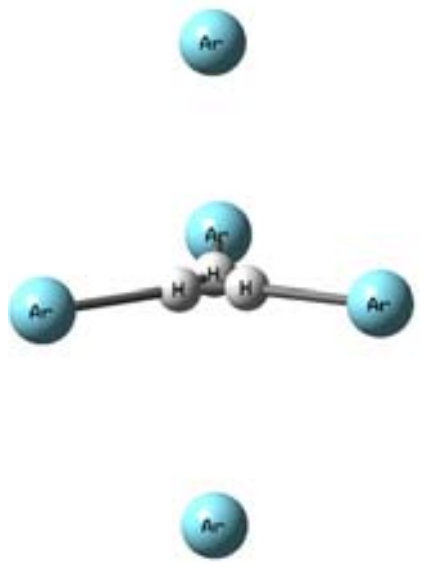

**$Ar_5H_3^+$**

Figure 1: Geometrical structures of $H_3^+$ and corresponding trapped noble gas clusters optimized at the B3LYP/6-311+G(d) level of theory.

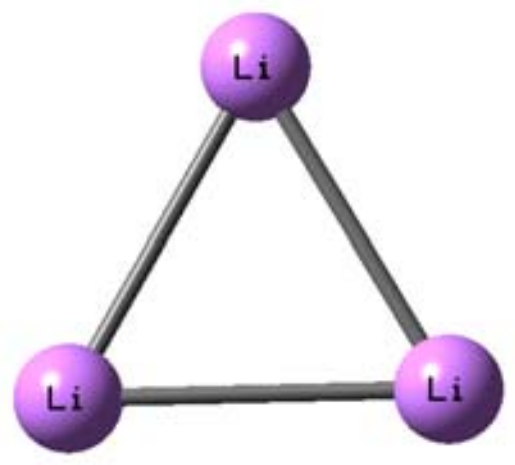

**$Li_3^+$**

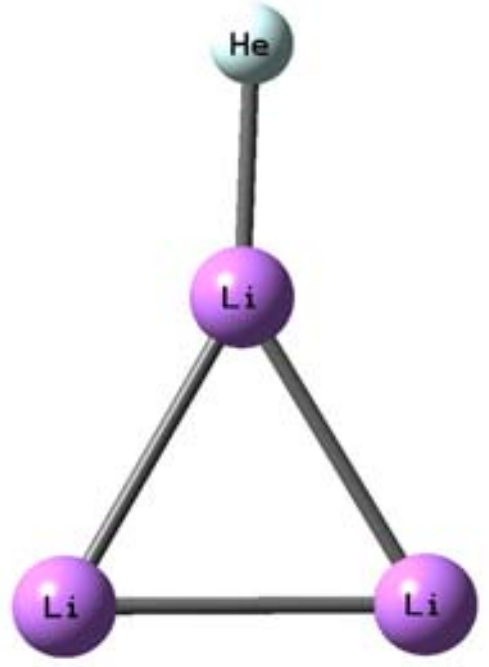

**$HeLi_3^+$**

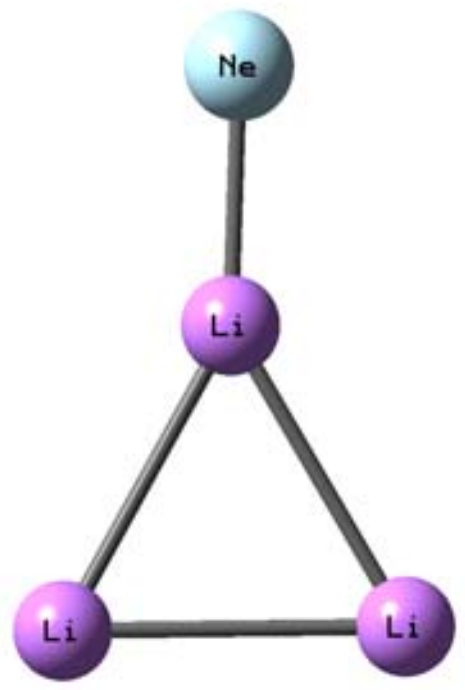

**$NeLi_3^+$**

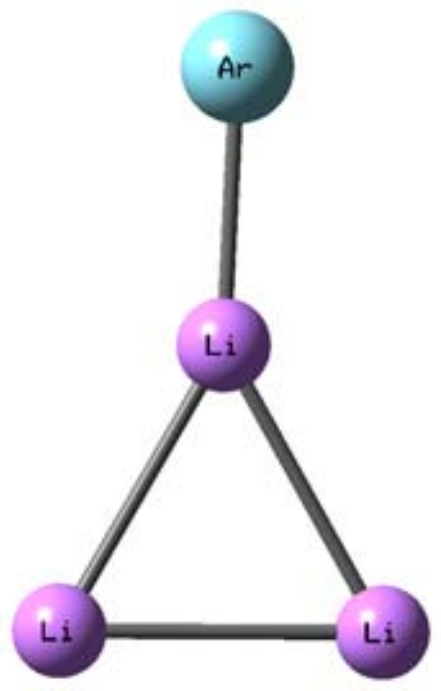

**$ArLi_3^+$**

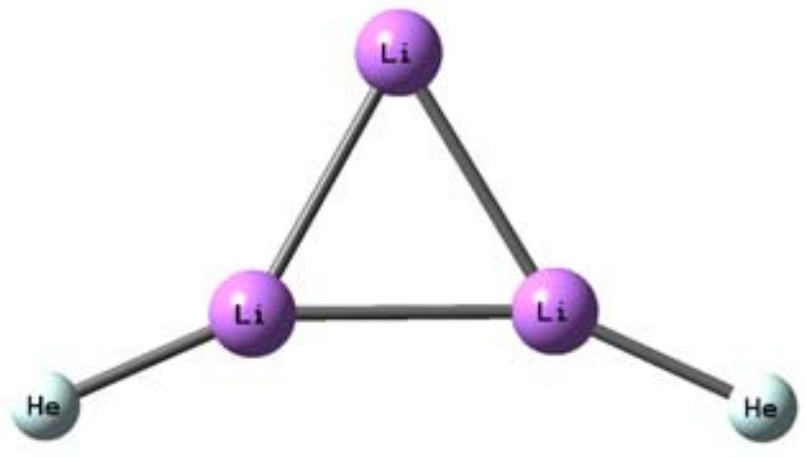

**$He_2Li_3^+$**

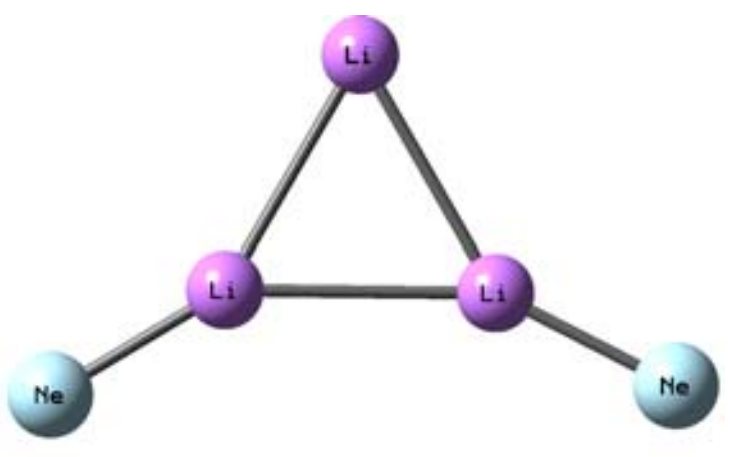

**$Ne_2Li_3^+$**

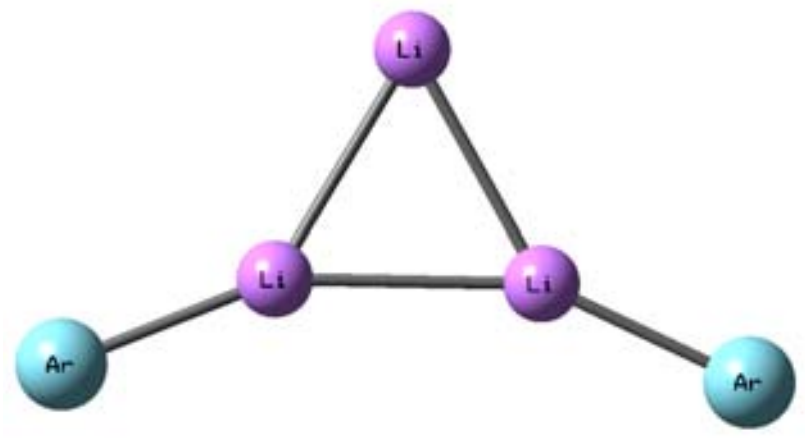

**$Ar_2Li_3^+$**

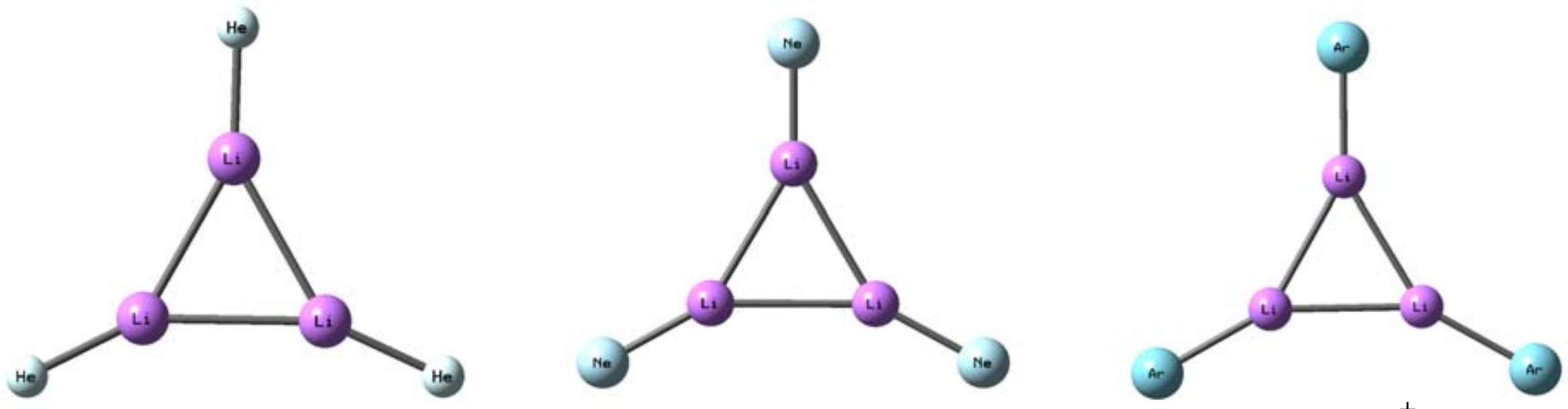

$He_3Li_3^+$ $Ne_3Li_3^+$ $Ar_3Li_3^+$

Figure 2: Geometrical structures of $Li_3^+$ and corresponding trapped noble gas clusters optimized at the B3LYP/6-311+G(d) level of theory.

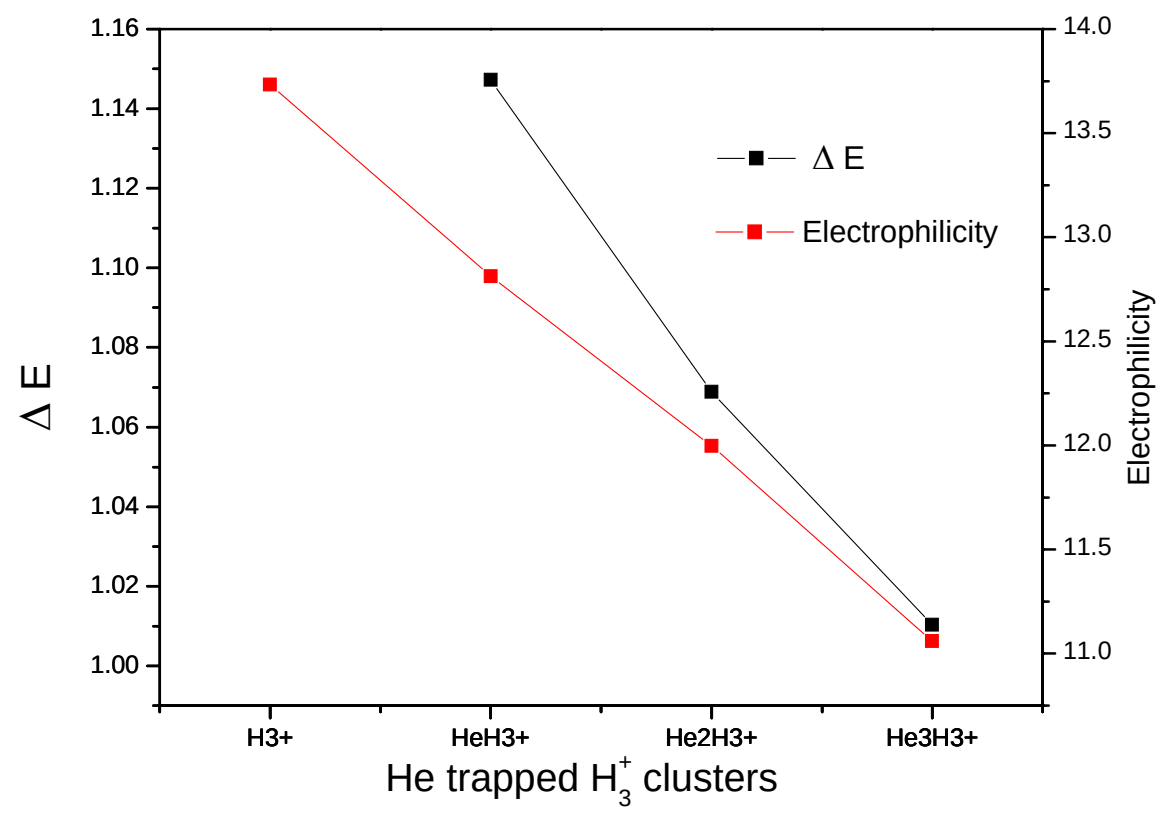

Figure 3a: Gain in energy for the formation of the complexes (ΔE, Kcal/mole) and electrophilicity (ω, eV) of the He-trapped $H_3^+$ clusters

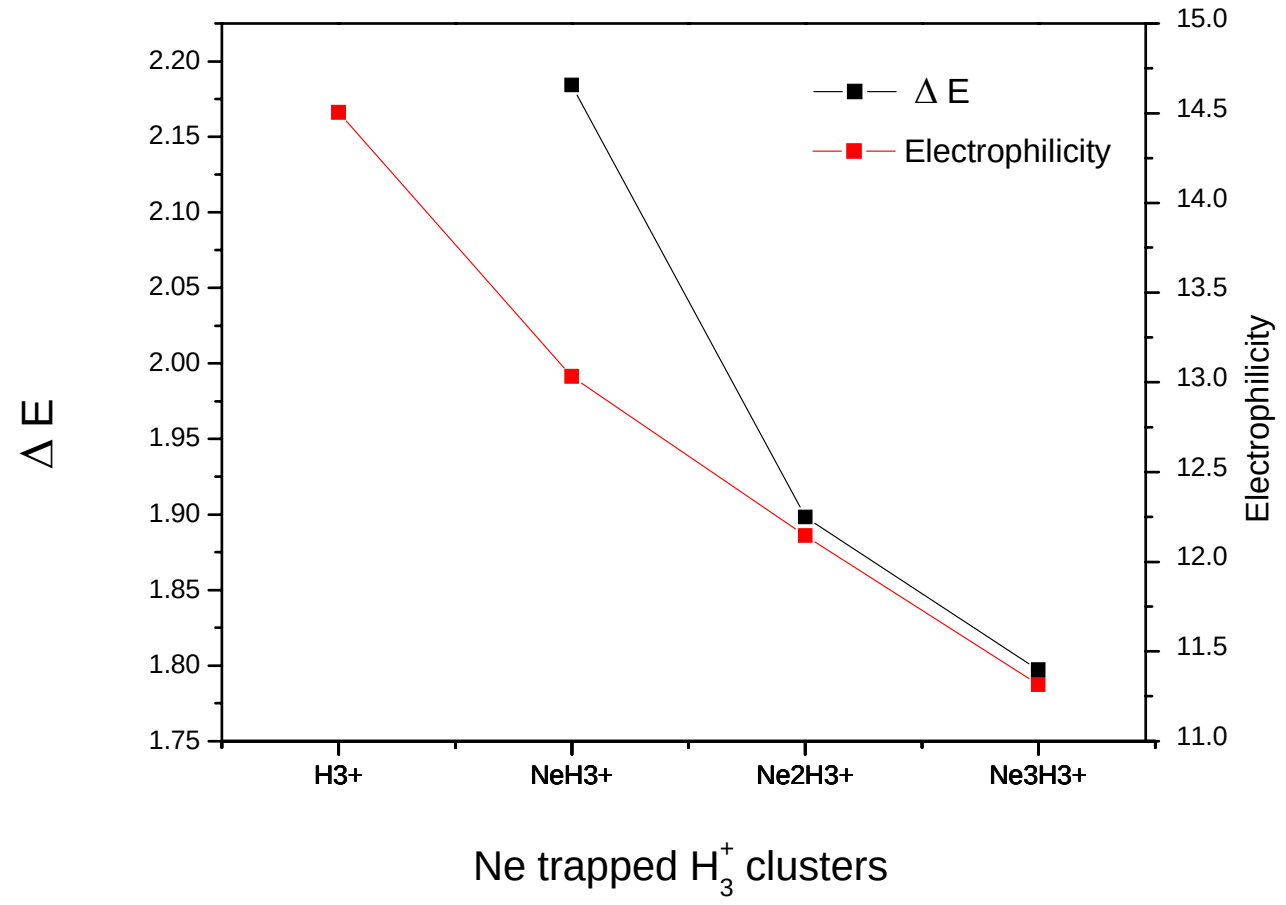

Figure 3b: Gain in energy for the formation of the complexes (ΔE, Kcal/mole) and electrophilicity (ω, eV) of the Ne-trapped $H_3^+$ clusters

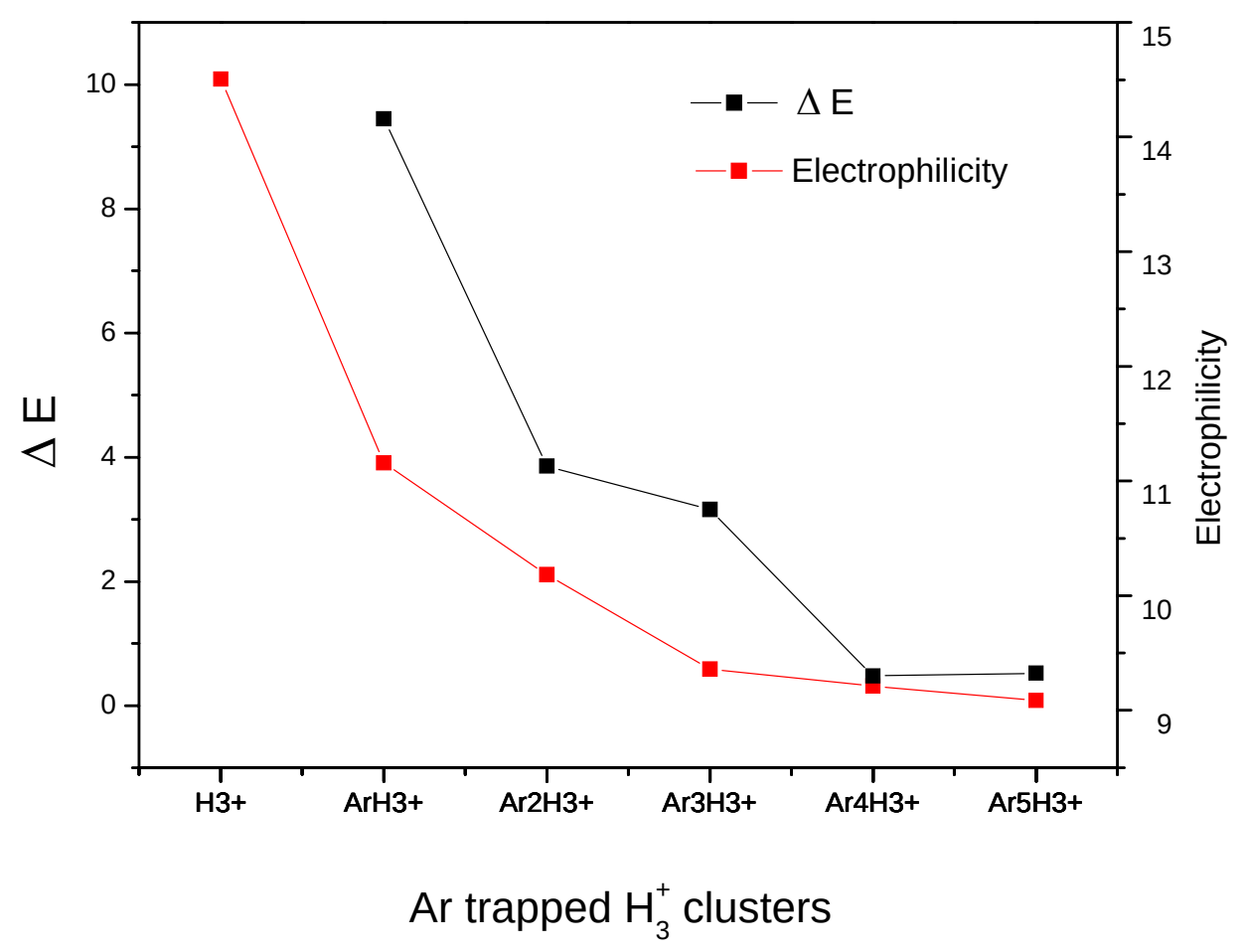

Figure 3c: Gain in energy for the formation of the complexes (ΔE, Kcal/mole) and electrophilicity (ω, eV) of the Ar-trapped $H_3^+$ clusters

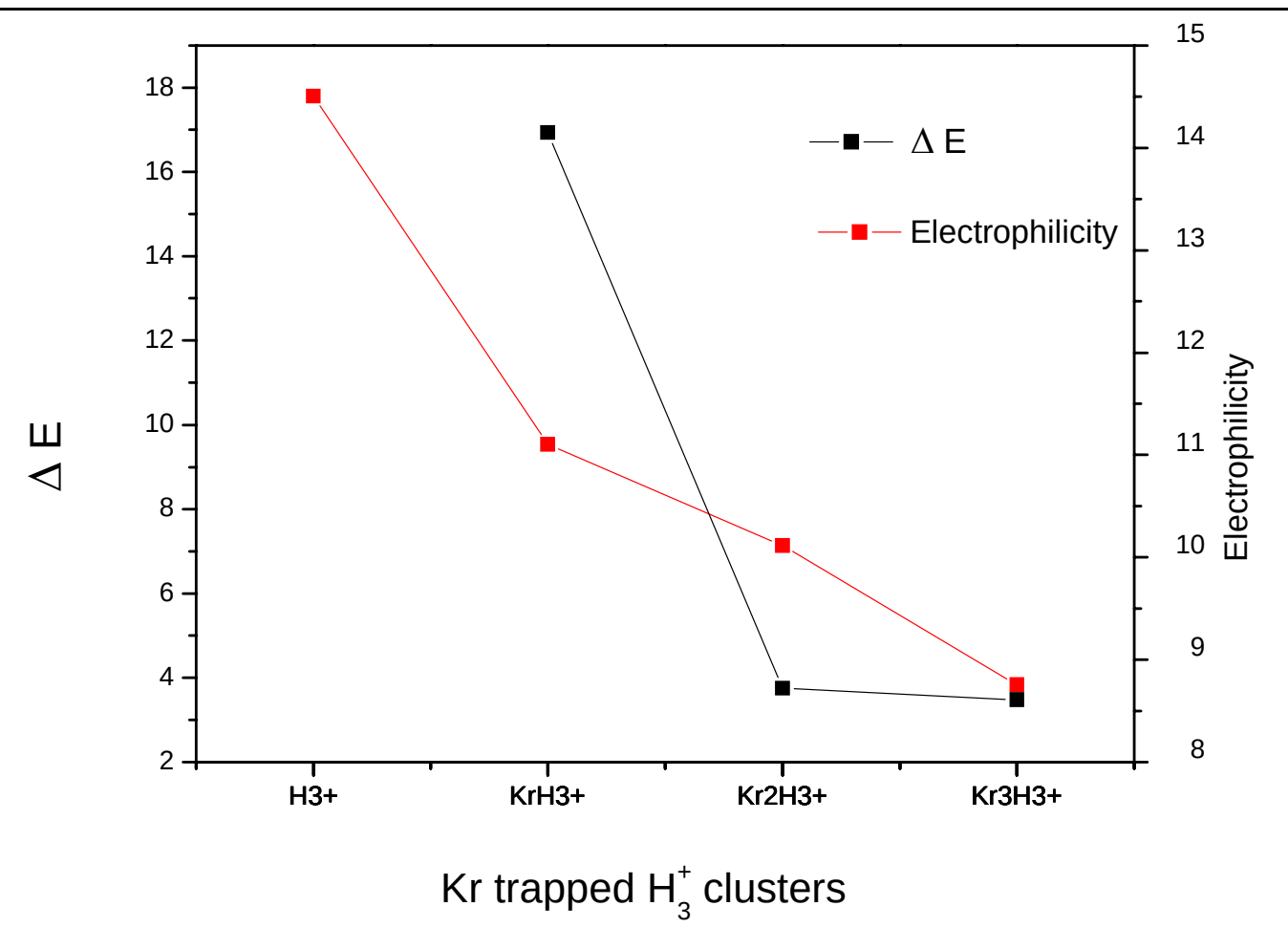

Figure 3d: Gain in energy for the formation of the complexes (ΔE, Kcal/mole) and electrophilicity (ω, eV) of the Kr-trapped $H_3^+$ clusters

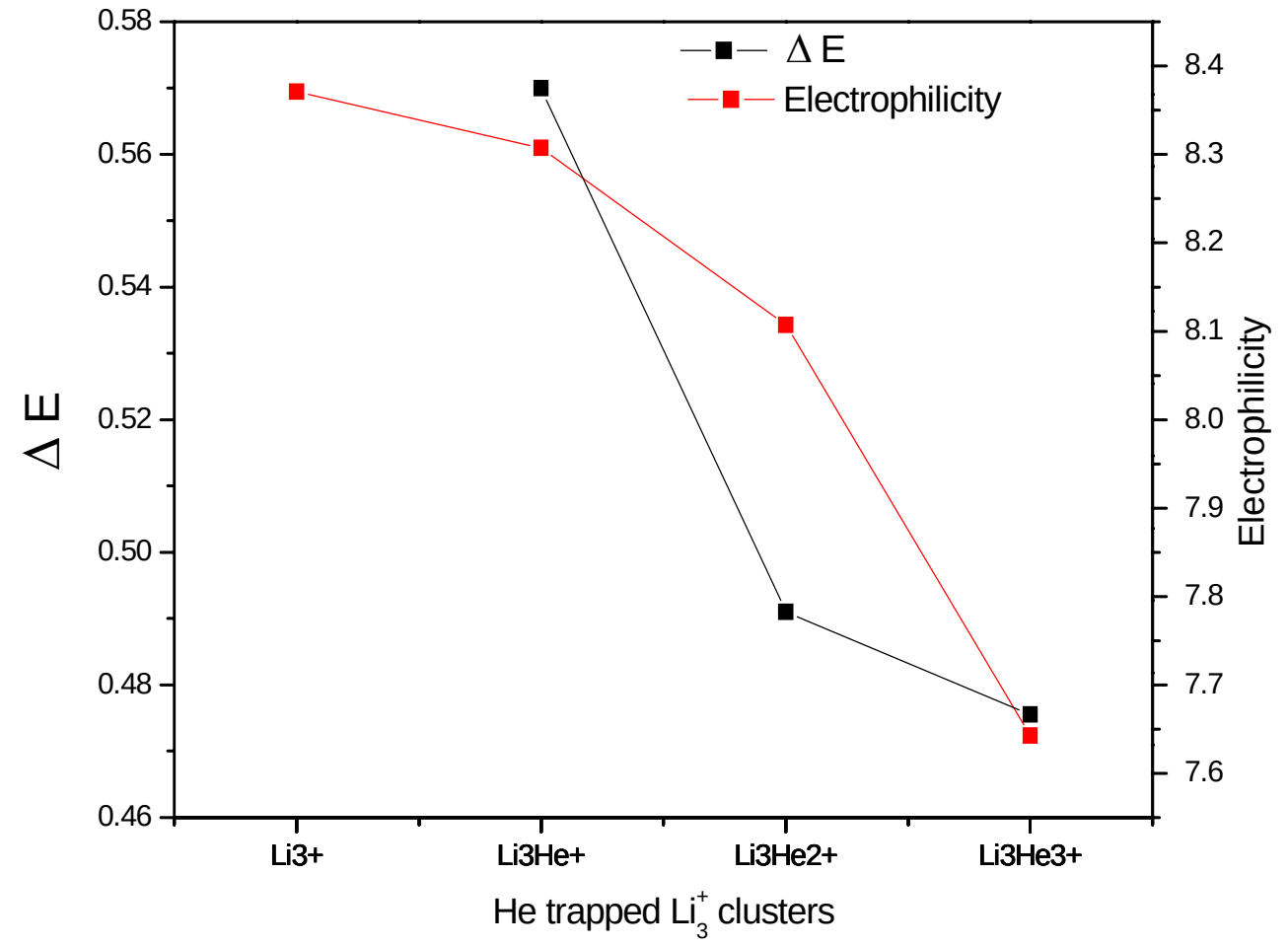

Figure 4a: Gain in energy for the formation of the complexes (ΔE, Kcal/mole) and electrophilicity (ω, eV) of the He-trapped $Li_3^+$ clusters

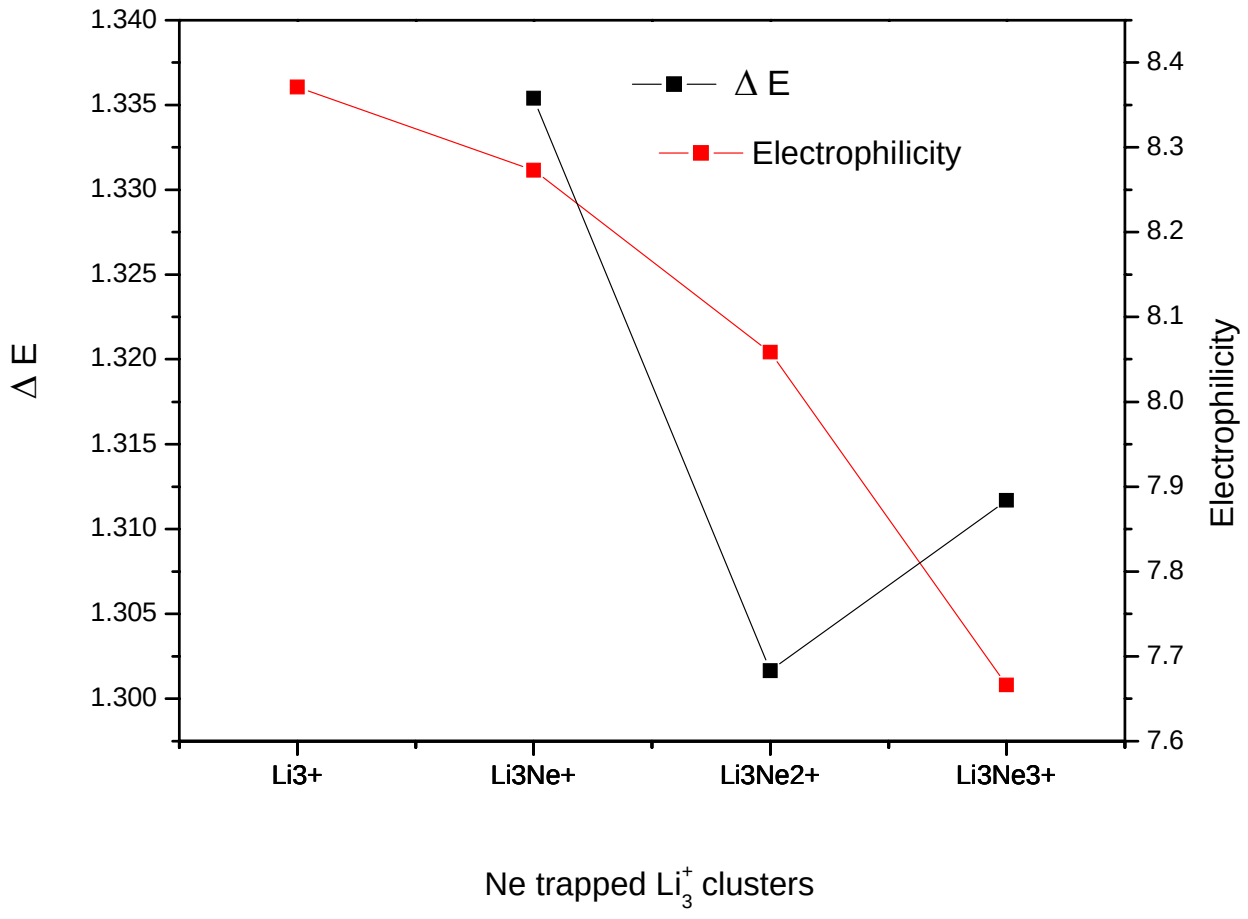

Figure 4b: Gain in energy for the formation of the complexes (ΔE, Kcal/mole) and electrophilicity (ω, eV) of the Ne-trapped $Li_3^+$ clusters

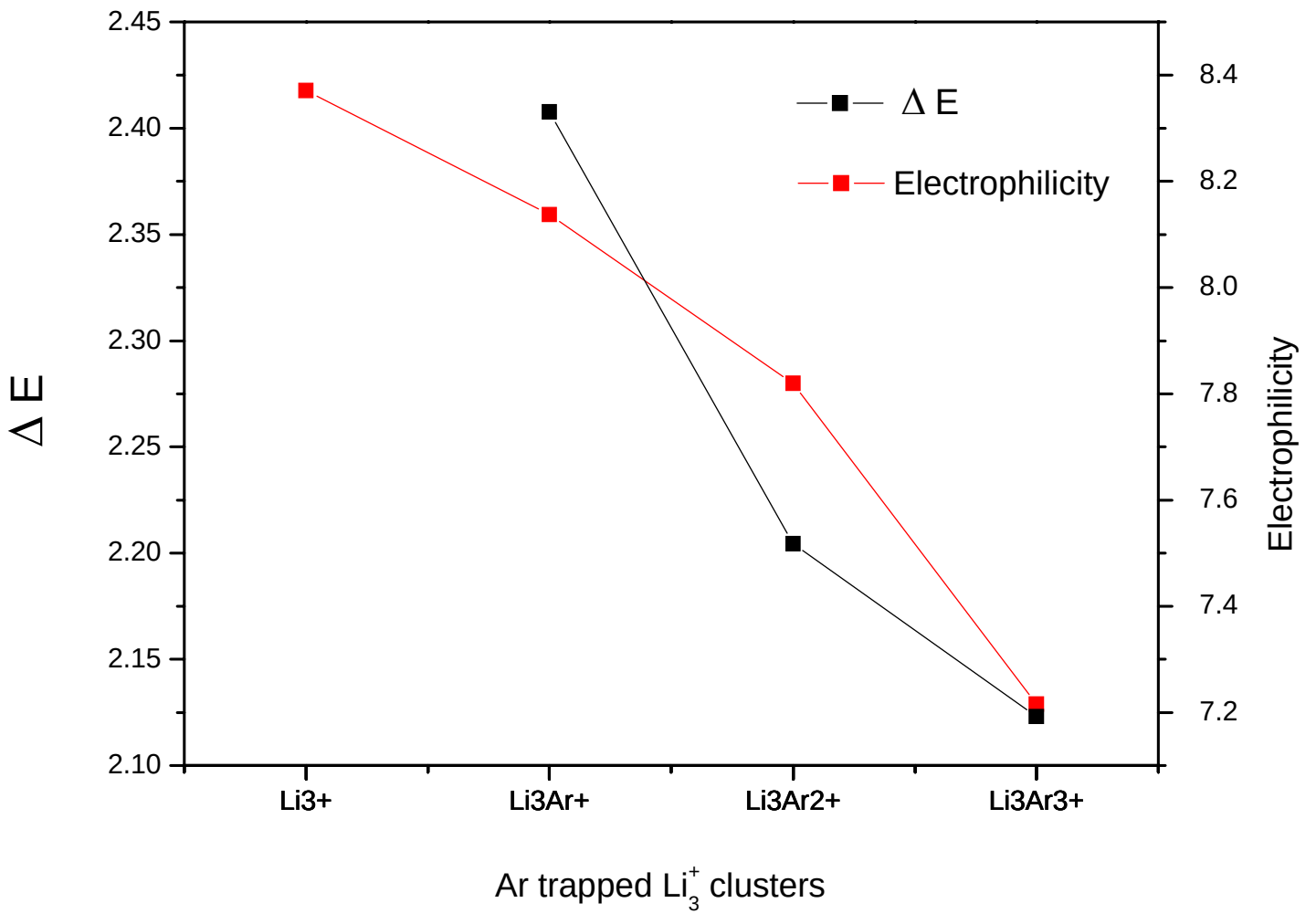

Figure 4c: Gain in energy for the formation of the complexes (ΔE, Kcal/mole) and electrophilicity (ω, eV) of the Ar-trapped $Li_3^+$ clusters

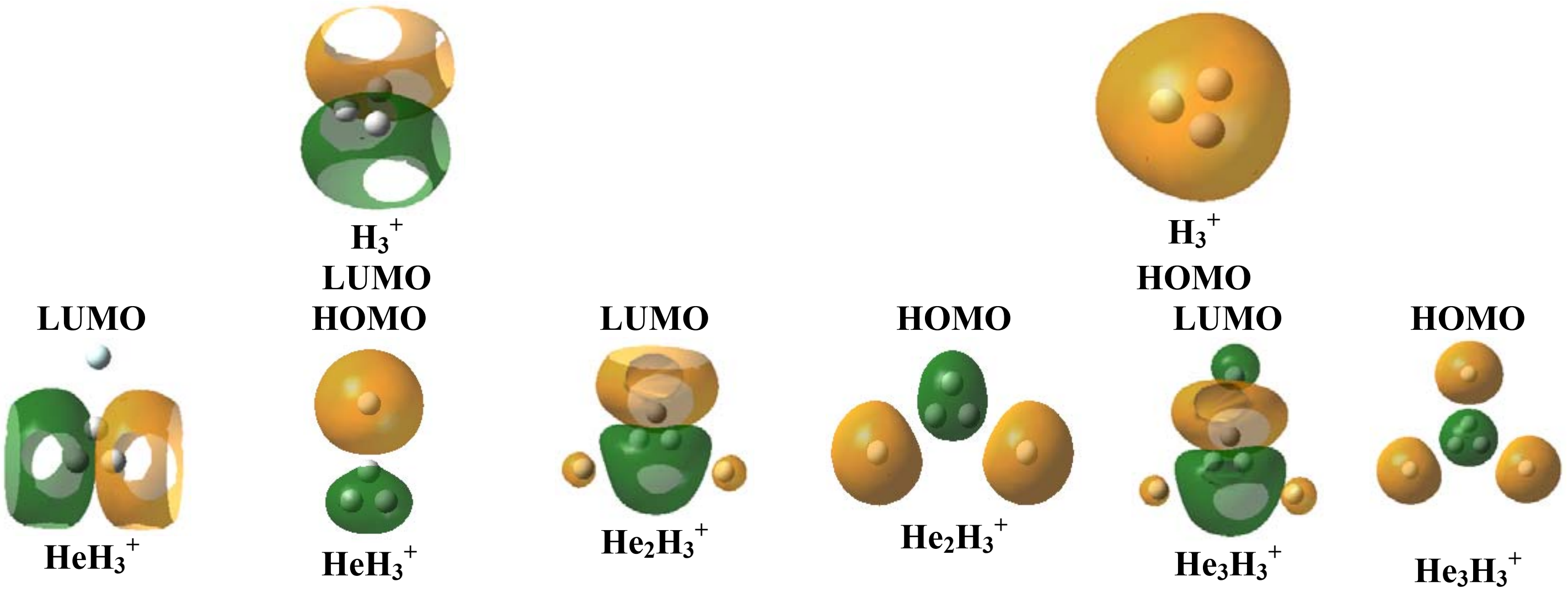

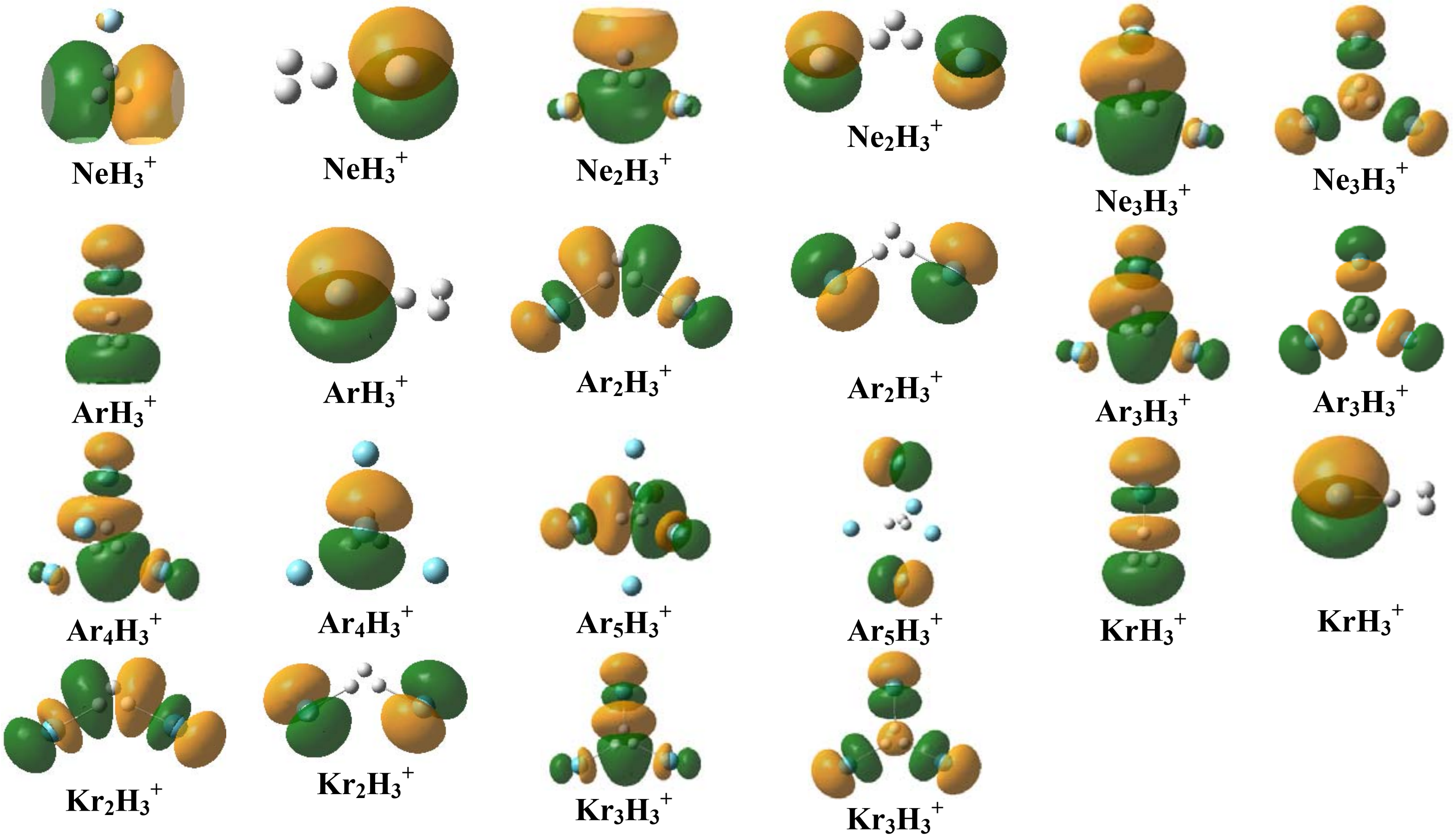

Figure 5: The frontier molecular orbitals (LUMO, HOMO) of $H_3^+$ and corresponding trapped noble gas clusters optimized at the B3LYP/6-311+G(d) level of theory.

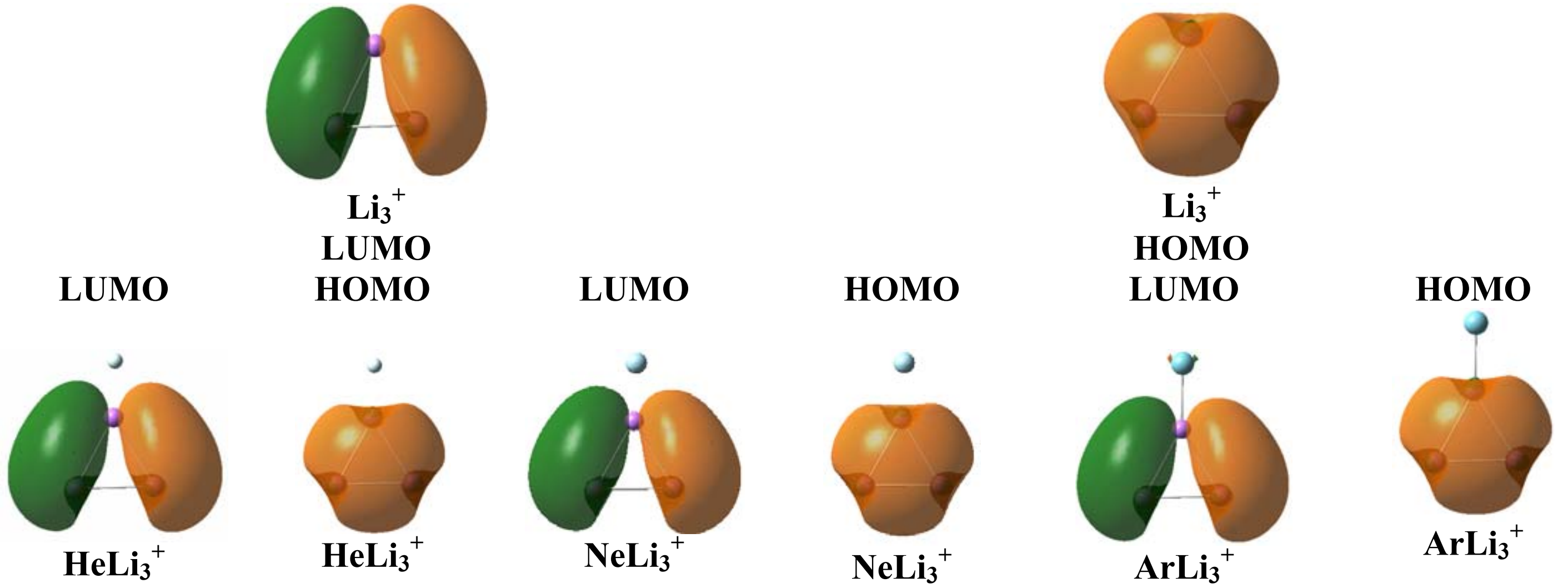

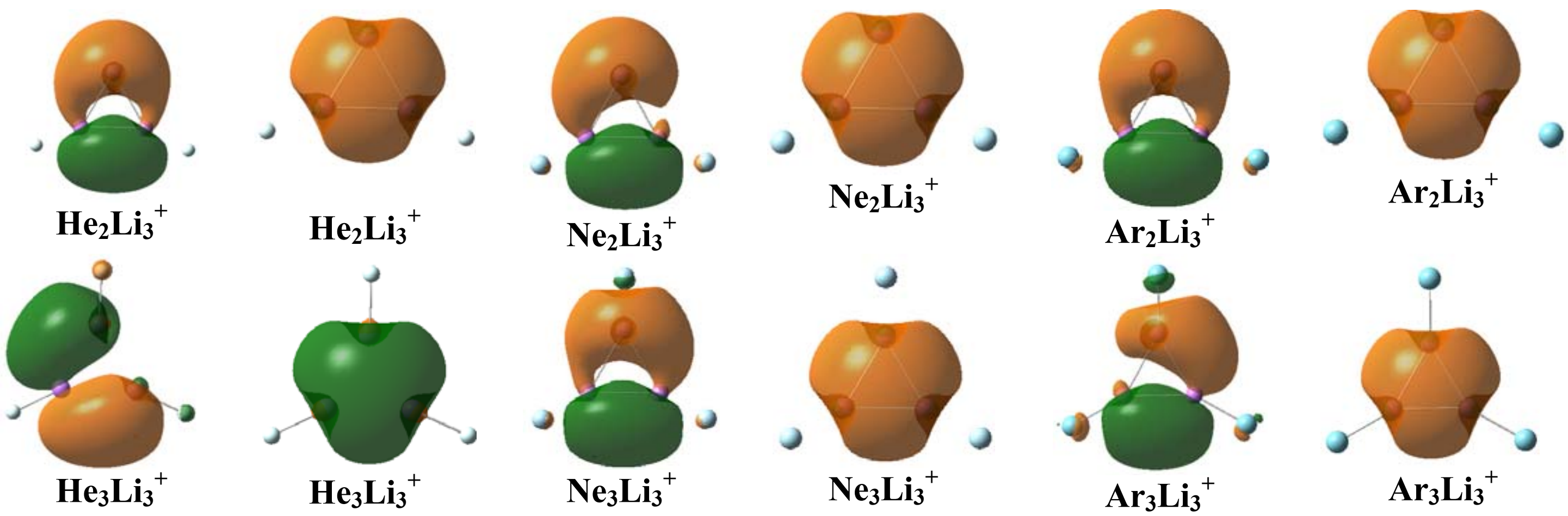

Figure 6: The frontier molecular orbitals (LUMO, HOMO) of $Li_3^+$ and corresponding trapped noble gas clusters optimized at the B3LYP/6-311+G(d) level of theory.